\documentclass[aps,prb,reprint,showpacs,superscriptaddress,longbibliography]{revtex4-2}
\usepackage{graphicx}

\makeatletter
\AtBeginDocument{\let\LS@rot\@undefined}
\makeatother

\usepackage{amsmath,amssymb,amsfonts,braket}
\usepackage{comment}
\usepackage{color}
\usepackage[T1]{fontenc}
\usepackage{textcomp}
\usepackage{hyperref}
\hypersetup{
setpagesize=false,
colorlinks=true,
linkcolor=blue,
citecolor=blue,
}
\usepackage{bm}
\usepackage{mathrsfs}
\usepackage{physics}
\usepackage{svg}
\usepackage{multirow}
\usepackage{pdfpages}
\usepackage{pgffor}

\newif\ifarXiv
\arXivfalse

\makeatletter
\let\MYcaption\@makecaption
\makeatother
\usepackage{subcaption}
\makeatletter
\let\@makecaption\MYcaption
\makeatother

\begin{document}

\preprint{APS/123-QED}

\title{
Quantum-geometry-controlled pseudogap and non-Fermi-liquid behavior
}

\author{Yu Miura}
\email{miura.yu.42c@st.kyoto-u.ac.jp}
\affiliation{Department of Physics, Graduate School of Science, Kyoto University, Kyoto 606-8502, Japan}

\author{Youichi Yanase}
\affiliation{Department of Physics, Graduate School of Science, Kyoto University, Kyoto 606-8502, Japan}

\date{\today}

\begin{abstract}
The quantum metric plays an essential role in flat-band superconductivity, governing the superfluid weight and coherence length. 
While previous studies have primarily focused on mean-field superconducting properties, the role of quantum geometry in fluctuation phenomena remains largely unexplored, despite the importance of fluctuations in flat-band systems. 
To address this issue, we employ a multiorbital T-matrix approximation that retains the orbital structure of the Bloch wave functions. 
By analyzing models in which the quantum metric can be tuned independently of the band dispersion, we show that an enhanced quantum metric suppresses the fluctuation-induced pseudogap in both flat-band and dispersive-band systems. 
In dispersive-band systems, this suppression is accompanied by a crossover from non-Fermi-liquid to Fermi-liquid behavior.
This spectral evolution suggests a quantum-metric-driven crossover from a BEC-like strong-coupling regime to a BCS-like weak-coupling regime.
More broadly, our results establish the quantum metric as a control parameter of interaction-driven fluctuation phenomena beyond mean-field theory.
\end{abstract}
\maketitle

\section{INTRODUCTION} 
Quantum geometry, encoded in Bloch wave functions, has emerged as a fundamental ingredient in electronic collective phenomena beyond conventional descriptions based on band energies~\cite{Liu2025-zk,GaoRMP2026,Kitamura2026-fs,Yu2025-xz}.
The geometric structure of Bloch states is characterized by the quantum geometric tensor, whose real and imaginary parts correspond to the quantum metric and the Berry curvature, respectively~\cite{Provost1980-gd}.
The Berry curvature underlies a wide range of topological and transport phenomena, including the quantum Hall effect and anomalous Hall effect~\cite{Xiao2010-or,Nagaosa2010-ye}.
More recently, the quantum metric has attracted growing attention as a distinct geometric quantity that governs physical responses, including nonlinear transport and superconductivity~\cite{Gao2023-nw,Wang2023,Peotta2015-tv,Liang2017-mk,Huhtinen2022-aq}.
The role of the quantum metric becomes particularly prominent in the flat-band limit, where contributions from group velocity are completely suppressed~\cite{Kitamura2026-fs}.

A striking manifestation of quantum geometry appears in flat-band superconductivity, where the quantum metric governs key superconducting properties despite the quenched electronic dispersion.
In a perfectly flat band, the conventional contribution to the superfluid weight vanishes with the band dispersion.
The quantum metric can instead generate a finite geometric contribution, thereby providing finite superfluid weight even in the flat-band limit~\cite{Peotta2015-tv,Liang2017-mk,Huhtinen2022-aq,Torma2022,Torma2023}.
More recently, it has also been shown that the quantum metric introduces a geometric length scale in superconductors, such as the superconducting coherence length, even in the flat-band limit~\cite{Chen2024-ye,Iskin2024-xl,Hu2025-dm}.
These results establish the quantum metric as a key ingredient of flat-band superconductivity at the mean-field level.

Beyond mean-field theories, quantum geometry has been shown to influence preformed-pair dynamics and finite-temperature superconducting properties in multiband and flat-band systems~\cite{Wang2020-tc,Hofmann2020-fp,Hofmann2023-ac}.
Nevertheless, how the quantum metric modifies the fluctuation-induced fermionic self-energy and the resulting  single-particle spectrum remains poorly understood.
The suppression of kinetic energy in flat-band systems enhances the relative importance of interactions, making pairing fluctuations particularly important~\cite{Hofmann2020-fp}.
Experimentally, tunneling spectroscopy of magic-angle twisted bilayer graphene has revealed a pseudogap that persists above the superconducting transition temperature~\cite{Oh2021-iz}, indicating anomalous normal-state spectral behavior beyond a simple mean-field BCS description.
It is therefore desirable to clarify how quantum geometry controls superconducting fluctuations and their signatures in the self-energy and single-particle spectrum.

To address this problem, we employ a multiorbital T-matrix approach together with models designed to control quantum geometry.
The T-matrix approximation provides a standard framework for incorporating pairing fluctuations across the BCS--BEC
crossover and for describing fluctuation-induced pseudogap behavior in the normal state~\cite{Nozires1985-br,Janko1997-he,Yanase1999-dw,Rohe2001-qg,Perali2002-yv,Tsuchiya2009-vi,YANASE20031}.
We retain the orbital structure of Bloch wave functions in the T-matrix formulation, thereby including the information that encodes quantum geometry of single-particle states~\cite{Kitamura2026-fs,Liang2017-mk}.
Because the roles of band dispersion and quantum geometry are generally intertwined in the multiorbital T-matrix calculation, we isolate the latter by varying the quantum metric while keeping the band dispersion fixed.

Using this framework, we find that the quantum metric suppresses the fluctuation-induced pseudogap and, in dispersive-band systems, drives a crossover from non-Fermi-liquid to Fermi-liquid behavior associated with the BCS--BEC crossover.
In the flat-band limit, the quantum metric strongly suppresses the pseudogap generated by superconducting fluctuations.
In systems with finite band dispersion, increasing the quantum metric likewise suppresses the pseudogap and changes the non-Fermi-liquid behavior of the low-energy self-energy to the Fermi-liquid behavior, signaling a continuous crossover from a BEC-like strong-coupling regime toward a BCS-like weak-coupling regime.
This crossover is accompanied by a quantum-metric-induced increase in the coherence length, which provides insight into its microscopic origin.

\section{MODEL}\label{sec:Model}
In this study, to formulate the effect of quantum geometry on many-body collective phenomena, we start from the Hamiltonian of a general two-band model with attractive interactions. 
The full Hamiltonian of the system is described using the wave vector $\bm{k}$, the orbital indices $l, l^{\prime}$, and the spin index $\sigma=\uparrow,\downarrow$ as follows: 
\begin{align}
    H = & \sum_{l,l'}\sum_{\bm{k}}\sum_\sigma \left[\hat{H}_0(\bm{k})\right]_{ll'}c_{l\sigma}^\dagger(\bm{k})c_{l'\sigma}(\bm{k})\notag\\
     & \hspace{0em}- \frac{1}{N}\sum_{l,l'} \sum_{\bm{k},\bm{k}',\bm{q}}\left[\hat{U}\right]_{ll'}c_{l\uparrow}^\dagger(\bm{k})c_{l'\downarrow}^\dagger(-\bm{k}+\bm{q})\notag\\
     &\hspace{9em} \times c_{l'\downarrow}(-\bm{k}'+\bm{q})c_{l\uparrow}(\bm{k}'),
\end{align}
where $\hat{H}_{0}(\bm{k})$ is the non-interacting Hamiltonian, $\hat{U}$ is the interaction matrix in the orbital space, and $N$ denotes the total number of unit cells.
We assume isotropic spin-singlet $s$-wave pairing. Since the spin degrees of freedom are trivial in the present formulation, the spin indices are suppressed hereafter.
In what follows, the interaction matrix is diagonal and proportional to the identity matrix, which is explicitly expressed as $\hat{U}=U\hat{I}$, with $U>0$ denoting the strength of the attractive interaction.  
We denote the normalized Bloch eigenstate of the $m$th band by $\ket{u_m(\bm{k})}$, satisfying $\hat{H}_0(\bm{k})\ket{u_m(\bm{k})}=\varepsilon_m(\bm{k})\ket{u_m(\bm{k})}$.
We further define $E_m(\bm{k})=\varepsilon_m(\bm{k})-\mu$, where $\mu$ is the chemical potential.
The orbital component of Bloch states is denoted by $u_l^m(\bm{k})=\braket{l}{u_m(\bm{k})}$.
The quantum metric of the $m$th band is defined as
\begin{align}
    g_{ij}^{m}(\bm{k})=\mathrm{Re}\left[\bra{\partial_iu_m}(1-\ket{u_m}\bra{u_m})\ket{\partial_ju_m}\right],
\end{align}
where $\ket{u_m}\equiv\ket{u_m(\bm{k})}$ and $\partial_i\equiv\partial/\partial k_i$.

To isolate the quantum-geometric effects in the multiorbital T-matrix calculation, we employ models in which the quantum metric can be varied independently of the band dispersion.
In some mean-field formulations, physical quantities such as the superfluid weight can be analytically decomposed into conventional and geometric contributions~\cite{Liang2017-mk,Peotta2015-tv}. 
In contrast, in the present many-body calculation, the effects of band dispersion and Bloch-wave-function geometry are intertwined, making such an analytical decomposition difficult.
However, by keeping the band dispersion fixed while varying the quantum metric, we can therefore identify changes arising from quantum geometry.

Accordingly, we consider two complementary two-band models with tunable quantum geometry.
As we introduce below, Model~1 consists of finite-range hopping terms and exhibits band-touching points, whereas Model~2 represents the isolated flat band without band touching and involves infinite-range hopping terms.
Comparing these qualitatively different models allows us to distinguish robust quantum-geometric effects from features specific to a particular band structure.

\subsection{Model 1}
We employ a variant of the two-band model introduced in Ref.~\cite{Oh2026}. This model has finite-range hopping and band-touching points and allows the quantum metric to be tuned by the parameter $0\leq d\leq1$ while keeping the band structure unchanged.
Specifically, the non-interacting Hamiltonian is expressed as $\hat{H}_0(\bm{k}) = \sum_{a=0,x,y,z} h_a(\bm{k}) \hat{\sigma}_a$, where $h_a(\bm{k})$ are given by:
\begin{align}
    h_0(\bm{k}) &= (2+\cos{2k_x}-\cos{2k_y})/2,\\[4pt]
    h_x(\bm{k}) &= d\sqrt{1-d^2}(1-\cos{2k_y}), \\[4pt]
    h_y(\bm{k}) &= 2d\cos{k_x}\sin{k_y}, \\[4pt]
    h_z(\bm{k}) &= (1-d^2)+ \{(2d^2-1)\cos{2k_y}+\cos{2k_x}\}/2.
\end{align}
The eigenenergies of the Hamiltonian are obtained as follows [Fig.~\ref{fig:bands_bare}(a)], 
\begin{align}
    \varepsilon_\pm(\bm{k})&=
                  \begin{cases}
                     2+\cos{2k_x}-\cos{2k_y}\\0
                  \end{cases},
\label{eq:dispersion_model1}
\end{align}
guaranteeing that the band structure is maintained regardless of the parameter $d$.
This model exhibits band-touching points, for example, at ${\bm k}=(\pm\pi/2,0)$.
Throughout the calculations for Model~1, we set the chemical potential to $\mu=0$; the chemical potential coincides with the energy of the lower band with flat band dispersion.

In contrast, the quantum metric of the flat band is obtained as a function proportional to $d^{2}$, as shown below:  
\begin{align}
    g_{xx}(\bm{k})&=d^2\frac{4\sin^2{k_x}\sin^2{k_y}}{(2+\cos{2k_x}-\cos{2k_y})^2},\\[4pt]
    g_{yy}(\bm{k})&=d^2\frac{4\cos^2{k_x}\cos^2{k_y}}{(2+\cos{2k_x}-\cos{2k_y})^2},\\[4pt]
    g_{xy}(\bm{k})&=-d^2\frac{\cos{2k_x}\sin{2k_y}}{(2+\cos{2k_x}-\cos{2k_y})^2}.
\end{align}
This ensures that the quantum metric of the system can be enhanced by increasing $d$ without changing the band structure. 

\subsection{Model 2}
\begin{figure}[htbp]
    \centering
    \includegraphics[width=1.0\linewidth]{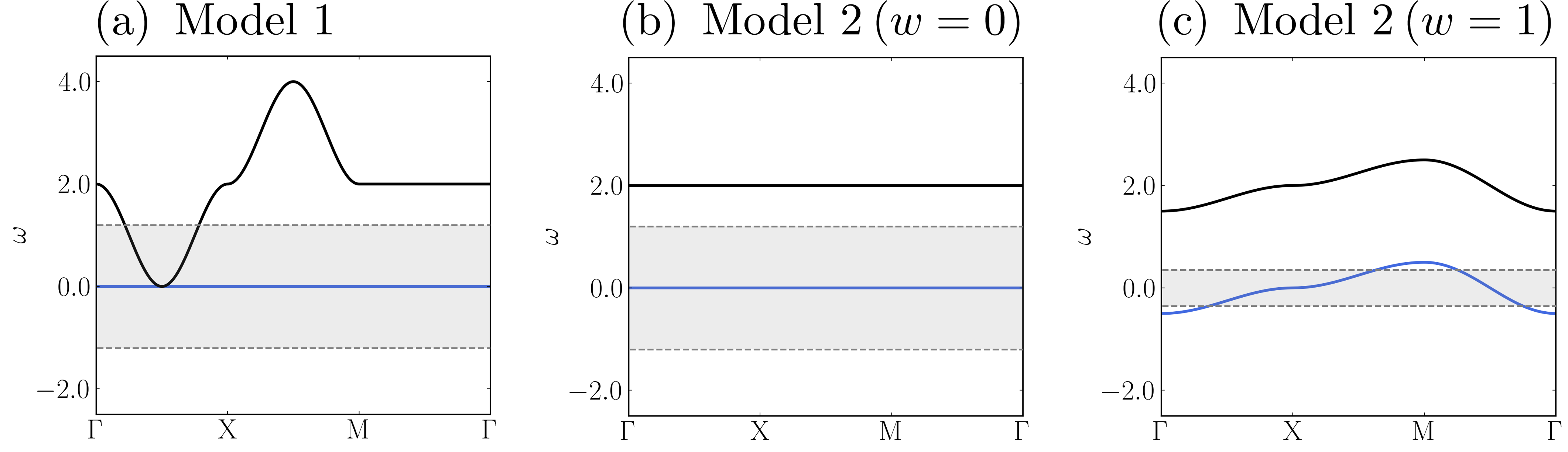}
    \caption{Band structures of the non-interacting Hamiltonian for (a) Model 1 and (b) Model 2 in the flat-band limit ($w=0$), and (c) Model 2 with finite band dispersion ($w=1$).
    The shaded regions indicate the energy window used in Figs.~\ref{fig:spectral_functions} and \ref{fig:crossover} for the spectral function in the interacting models. 
    }
    \label{fig:bands_bare}
\end{figure}

Model~2 is a modified version of a two-band model introduced in Refs.~\cite{Hofmann2022-tc,Hofmann2023-ac}.
It realizes isolated flat bands without band touching and allows the quantum metric to be tuned via the parameter \(\chi\) while keeping the band dispersion unchanged.
Specifically, the model is described by the following non-interacting Hamiltonian:
\begin{align}
    \hat{H}_0(\bm{k}) = 
    \begingroup
    \renewcommand{\arraystretch}{1.5}
    \begin{pmatrix}
        0 & -ie^{-i\alpha(\bm{k})} \\
        ie^{i\alpha(\bm{k})} & 0
    \end{pmatrix}
    \endgroup
    .
\end{align} 
where
\begin{align}
    \alpha(\bm{k}) = \chi (\sin k_x + \sin k_y).
\end{align}
This model may be regarded as a theoretical toy model because it contains infinite-range hopping terms in real space. 
The eigenenergies of this Hamiltonian are 
\begin{align}
    \varepsilon_\pm(\bm{k})=\pm1,
\end{align}
demonstrating two isolated flat bands without band touching [Fig.~\ref{fig:bands_bare}(b)], unlike Model~1.
We set the chemical potential to $\mu=-1$, which corresponds to the energy of the lower flat band.
The same chemical potential is also used when the finite band dispersion is introduced in the following.
The quantum metric of this system is obtained as follows, yielding a function proportional to $\chi^{2}$: 
\begin{align}
    g_{ij}(\bm{k})=\chi^2\cos{k_i}\cos{k_j}/4\quad(i,j=x,y).
\end{align}

Comparing these two models allows us to test whether the pseudogap controlled by the quantum metric, which we demonstrate below, is robust against qualitative differences in the models such as the band touching.
 
We also study a model defined by introducing finite band dispersion into Model~2.
Specifically, a term $\delta(\bm{k})$ defined by the following equation is added to the diagonal part of the original Hamiltonian: 
\begin{align}
    \delta(\bm{k}) &= - \frac{w}{4}(\cos{k_x}+\cos{k_y}).
\end{align} 
With this term included, the non-interacting Hamiltonian takes the form
\begin{align}
    \hat{H}_0(\bm{k}) = 
    \begingroup
    \renewcommand{\arraystretch}{1.5}
    \begin{pmatrix}
        \delta(\bm{k}) & -ie^{-i\alpha(\bm{k})} \\
        ie^{i\alpha(\bm{k})} & \delta(\bm{k})
    \end{pmatrix}
    \endgroup
    .
\end{align} 
This term continuously transforms the bands from flat bands to weakly dispersive bands [Fig.~\ref{fig:bands_bare}(c)],
\begin{align}
     \varepsilon_\pm(\bm{k})=\pm 1 - \frac{w}{4}(\cos{k_x}+\cos{k_y}).
\end{align}
The parameter $w$ exactly corresponds to the bandwidth.
Importantly, even though the band energies are altered by introducing dispersion in this manner, the quantum metric of the system is preserved in its original form and can still be controlled by the parameter $\chi$.
By systematically varying the model parameters, we can not only identify phenomena unique to flat-band systems through comparative analysis but also clarify quantum-geometry-driven crossovers in dispersive-band systems.

\section{METHOD}\label{sec:Method}
Throughout this paper, we focus on the normal state above the superconducting transition temperature $T_{\rm c}$.
To microscopically incorporate the superconducting fluctuations and quantum-geometric effects, we formulate our theory based on the T-matrix approximation (TMA)~\cite{Yanase1999-dw,Rohe2001-qg,Perali2002-yv,Tsuchiya2009-vi,YANASE20031} while retaining the orbital degrees of freedom of the wave functions in a matrix form. 
In general, the effect of quantum geometry in a multiband system originates from the nontrivial geometric structure in momentum space of the Bloch wave functions, which consist of a superposition of multiple orbital components. 
Therefore, in the framework of this study, which microscopically treats the interplay between quantum geometry and many-body effects, it is essential to explicitly take into account the orbital degrees of freedom $l, l^{\prime}$ and treat various physical quantities as matrices in the orbital space.

First, we define the Matsubara Green's function for free electrons and the bare particle-particle correlation function, which are the fundamental building blocks of the Feynman diagrams, as matrices in the orbital space. 
Using the eigenvalues and orbital components $u_l^m(\bm{k})$ defined above, the free-electron Green's function is defined as follows:
\begin{align}
    \left[\hat{G}_0(\bm{k}, i\omega_n)\right]_{ll'} &= \left[ \frac{\hat{I}}{(i\omega_n + \mu) \hat{I} - \hat{H}_0(\bm{k})} \right]_{ll'}\\[7pt]
    &=\sum_m\frac{1}{i\omega_n-E_m(\bm{k})}u_l^m(\bm{k})u_{l'}^m(\bm{k})^*.
\end{align}
Subsequently, the particle-particle correlation function is defined as:
\begin{align}
    &\left[\hat{\Pi}(\bm{q},i\nu_n)\right]_{ll'} \notag\\[5pt]
    =&\frac{1}{N\beta}\sum_{\bm{k},\omega_n}\left[\hat{G}_0(\bm{k},i\omega_n)\right]_{ll'}
    \left[\hat{G}_0(\bm{-k+q},-i\omega_n+i\nu_n)\right]_{ll'}\\[7pt]
    =&\frac{1}{N\beta}\sum_{\bm{k},\omega_n}\sum_{m,m'}\frac{1}{i\omega_n-E_m(\bm{k})}\frac{1}{i\nu_n-i\omega_n-E_{m'}(\bm{-k+q})}\notag\\
    &\hspace{1.5em}\times u_l^m(\bm{k})u_{l'}^m(\bm{k})^*u_l^{m'}(\bm{-k+q})u_{l'}^{m'}(\bm{-k+q})^*\\[7pt]
    =&-\frac{1}{N}\sum_{\bm{k}}\sum_{m,m'}\frac{1-f(E_{m}(\bm{k}))-f(E_{m'}(\bm{-k+q}))}{i\nu_n-E_{m}(\bm{k})-E_{m'}(\bm{-k+q})}\notag\\
    &\hspace{1.5em}\times{u^{m}_{l}(\bm{k})}{u^{m}_{l'}(\bm{k})}^*{u^{m'}_{l}(\bm{-k+q})}{u^{m'}_{l'}(\bm{-k+q})}^*.
\end{align}

In this formulation, the quantum geometric properties of the system are seamlessly incorporated through the orbital components of the wave functions $u_{l}^{m}(\bm{k})$, and they are connected to the quantum metric later. 
Using the particle-particle correlation function defined above, we include multiple scattering processes mediated by the attractive interaction to infinite order, treating this as a microscopic description of the superconducting fluctuations.
By summing the multiple particle-particle scattering processes, we derive the T-matrix, which directly reflects the quantum-geometric effects, and calculate its feedback on the single-particle spectrum. 
Specifically, by calculating the sum of ladder diagrams using the interaction matrix and the particle-particle correlation function, we analytically obtain the T-matrix that describes the superconducting fluctuations as follows:
\begin{align}
    \hat{\Gamma}(\bm{q},i\nu_n)&=-\hat{U}-\hat{U}\hat{\Pi}(\bm{q},i\nu_n)\hat{U}\notag\\[4pt]
    &\hspace{4.0em}-\hat{U}\hat{\Pi}(\bm{q},i\nu_n)\hat{U}\hat{\Pi}(\bm{q},i\nu_n)\hat{U}-\cdots\\
    &=-\frac{\hat{I}}{\hat{{I}}-\hat{U}\hat{\Pi}(\bm{q},i\nu_n)}\hat{U}.
\end{align}
Here, each scattering process is represented as a matrix product, which accurately describes successive transitions between Cooper pairs of various orbitals in the intermediate states. 

The superconducting transition occurs when the particle-particle scattering matrix develops a pole at zero center-of-mass momentum and zero bosonic Matsubara frequency, known as the Thouless criterion~\cite{Thouless1960-yp}.
In the present multiorbital formulation, we characterize this instability by the lowest eigenvalue of the T-matrix, defined as:
\begin{align}
    \Gamma_{\rm min}(\bm{q})=\min\left[\text{spec}\left(\hat{\Gamma}(\bm{q}, 0)\right)\right].
\end{align}
Consequently, the transition temperature $T_{\rm c}$ is determined as the highest temperature at which
\begin{align}
    \Gamma_{\rm min}(\bm{0})\xrightarrow[T\rightarrow T_{\rm c}]{}-\infty.
\end{align}
We note that the transition temperature determined by the Thouless criterion within the present T-matrix approximation should be distinguished from the Berezinskii–-Kosterlitz–-Thouless transition temperature in strictly two-dimensional systems. 
Here, we use this temperature as a characteristic pairing-instability scale to investigate superconducting fluctuation effects.
The dominant component of the T-matrix $\Gamma_{\text{min}}(\bm{q})$ is plotted to represent the superconducting fluctuations in subsequent discussions.

Next, to evaluate the impact of superconducting fluctuations on the single-particle spectrum, we calculate the TMA self-energy~\cite{Perali2002-yv,Tsuchiya2009-vi} by convoluting the T-matrix with the free-hole Green's function as follows:
\begin{align}
    &\left[\hat{\Sigma}_{\rm TMA}(\bm{k},i\omega_n)\right]_{ll'}\notag\\[5pt]
    &\hspace{1em}=-\frac{1}{N\beta}\sum_{\bm{q},\nu_n}\left[\hat{\Gamma}(\bm{q},i\nu_n)\right]_{ll'}\notag\\
    &\hspace{6em}\times\left[\hat{G}_0^h(\bm{k-q},i\omega_n-i\nu_n)\right]_{ll'}e^{i\nu_n0^+},
\end{align}
where $\hat{G}_{0}^{h}$ represents the free-hole Matsubara Green's function, defined as
\begin{align}
    \left[\hat{G}_0^h(\bm{k},i\omega_n)\right]_{ll'}=-\left[\hat{G}_0(-\bm{k},-i\omega_n)\right]_{l'l}.
\end{align}

Finally, we incorporate the multiorbital self-energy matrix into the Dyson equation. 
In our numerical calculations, we first perform the analytic continuation of the TMA self-energy from the Matsubara frequencies to the real-frequency axis using the Pad\'e approximation~\cite{Vidberg1977-rw}, yielding the retarded self-energy $\hat{\Sigma}^{\rm R}(\bm{k}, \omega) = \hat{\Sigma}_{\rm TMA}(\bm{k}, i\omega_n \to \omega + i\eta)$, where $\eta$ is a positive infinitesimal. 
Using the retarded self-energy, we obtain the retarded full Green's function:
\begin{align}
    \hat{G}^{\rm R}(\bm{k}, \omega)=\frac{\hat{I}}{(\omega+\mu+i\eta)\hat{I}-\hat{H}_0(\bm{k})-\hat{\Sigma}^{\rm R}(\bm{k}, \omega)}.
\end{align}
Subsequently, the single-particle spectral function $A(\bm{k},\omega)$, which reflects many-body scattering due to superconducting fluctuations, is calculated by taking the imaginary part of the trace over the orbital degrees of freedom:
\begin{align}
    A(\bm{k},\omega)=-\frac{1}{\pi}{\rm Im\,Tr}\,\hat{G}^{\rm R}(\bm{k}, \omega).
\end{align}
This sequence of formulations bridges the microscopic quantum-geometric effects embedded in the multiorbital superconducting fluctuations to the single-particle spectrum, allowing us to demonstrate the pseudogap and non-Fermi-liquid behaviors in the following section.

\section{RESULTS}\label{sec:Results}
\begin{figure*}[htbp]
    \centering
    \includegraphics[width=1.0\linewidth]{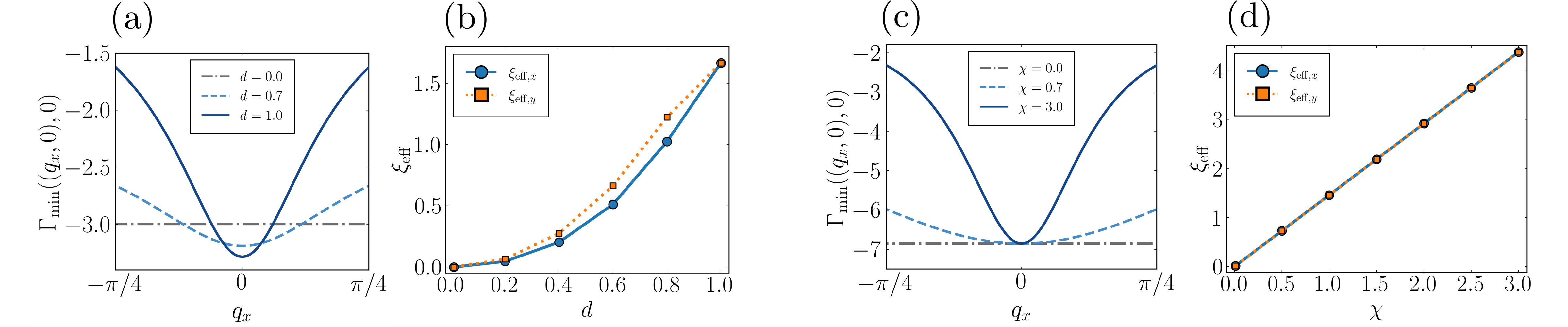}
    \caption{Momentum dependence of the lowest eigenvalue $\Gamma_{\rm min}(\bm{q})$ of the T-matrix for (a) Model~1 and (c) Model~2 in the flat-band limit ($w=0$). We plot along the $q_x$ direction at $q_y=0$. 
    Effective coherence length $\xi_{\rm eff}$ as a function of the quantum metric parameter is shown for (b) Model~1 and (d) Model~2. 
    Because of the absence (presence) of four-fold rotation symmetry, $\xi_{{\rm eff},x} \ne \xi_{{\rm eff},y}$ ($\xi_{{\rm eff},x} = \xi_{{\rm eff},y}$) in Model~1 (Model~2).
    The interaction parameter is set to $U = 0.5$ for Model~1 [(a), (b)] and $U = 1.0$ for Model~2 [(c), (d)]. 
    All results are evaluated at a temperature $T=1.2T_{\mathrm c}$.}
    \label{fig:T_matrix_flat}
\end{figure*}
\begin{figure*}[htbp]

    \centering
    \includegraphics[width=1.0\linewidth]{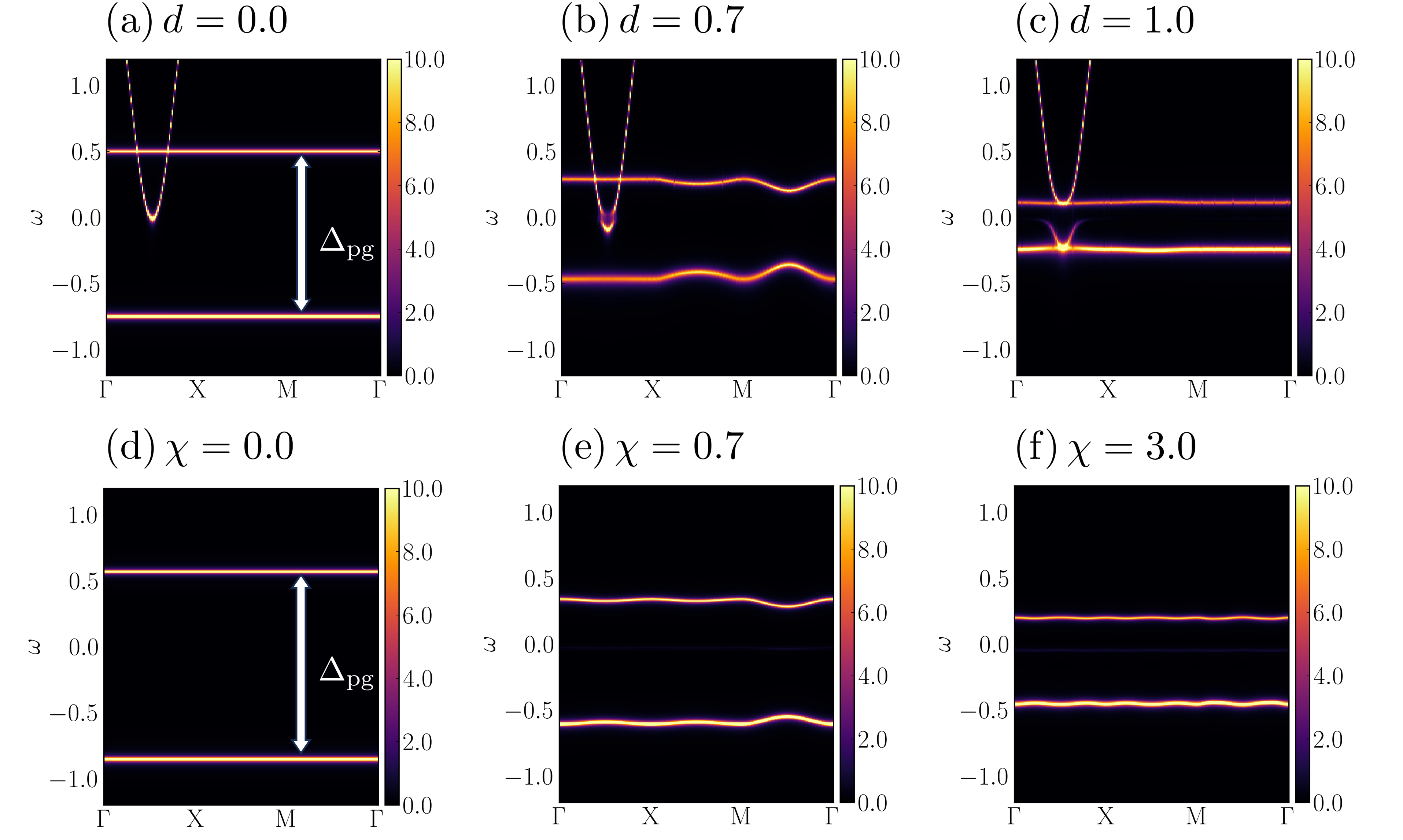}
    \caption{Single-particle spectrum in the low-energy region for various quantum metric parameters in the flat-band models. 
    Intensity plots for Model~1 with (a) $d=0.0$, (b) $d=0.7$, (c) $d=1.0$, and for Model~2 with (d) $\chi=0.0$, (e) $\chi=0.7$, (f) $\chi=3.0$. 
    The interaction parameter is set to $U = 0.5$ for Model~1 [(a)--(c)] and $U = 1.0$ for Model~2 [(d)--(f)].
    All results are evaluated at $T=1.2T_{\mathrm c}$.}
    \label{fig:spectral_functions}
\end{figure*}
In this section, we demonstrate that quantum geometry acts as a driving force for the suppression of the pseudogap phenomena and the crossover from non-Fermi-liquid to Fermi-liquid behavior.
In what follows, we first analyze the fluctuation effects in the flat-band limit, followed by an examination of dispersive-band systems.
Throughout this section, all calculations are performed in the normal phase slightly above the superconducting transition temperature ($T = 1.2T_{\mathrm{c}}$), where $T_{\mathrm{c}}$ is determined separately for the parameter of each model, namely $d$ or $\chi$.
At fixed interaction strength $U$, the transition temperature $T_{\mathrm{c}}$ depends on $d$ in Model~1, whereas it is independent of $\chi$ in Model~2.

\subsection{Quantum-geometry-controlled pseudogap in flat-band systems}\label{subsec:pg_flat}
To clarify the superconducting fluctuations in the flat-band limit ($w=0$), we first calculate the lowest eigenvalue of the T-matrix $\Gamma_{\rm min}(\bm{q})$ for both Model~1 and Model~2 while varying the quantum metric parameters, $d$ and $\chi$. 
Figures~\ref{fig:T_matrix_flat}(a) and \ref{fig:T_matrix_flat}(c) show the momentum dependence of $\Gamma_{\rm min}(\bm{q})$ for Model~1 and Model~2, respectively.
In both models, $\Gamma_{\rm min}(\bm{q})$ does not exhibit momentum dependence in the absence of quantum-geometric effects ($d=0$ or $\chi=0$), indicating spatially local fluctuations.
However, the negative peak of $\Gamma_{\rm min}(\bm{q})$ appears when the quantum geometry is switched on, and the peak becomes sharper as the quantum metric increases. 
These results demonstrate that the long-range superconducting fluctuations develop with the quantum metric, consistent with the fact that the coherence length of flat-band superconductors is given by the quantum metric~\cite{Chen2024-ye,Hu2025-dm}.
To quantify the spatial extent of the superconducting fluctuations, we introduce an effective coherence length $\xi_{\mathrm{eff}}$ from the momentum-space width of the T-matrix.
Specifically, it is evaluated from the dispersion of $\Gamma_{\rm min}(\bm{q})$ around $\boldsymbol{q} = \mathbf{0}$ as follows:
\begin{align}
    (\xi_{\rm eff}^2)_{ij} = \frac{1}{2\Gamma^{-1}_{\rm min}(0)}\left.\pdv{\Gamma^{-1}_{\rm min}(\bm{q})}{q_i}{q_j}\right|_{\bm{q}\to0}.
\end{align}
The directional coherence lengths are defined as
\begin{align}
    \xi_{{\rm eff},i}=\sqrt{(\xi_{\rm eff}^2)_{ii}},\quad i=x,y.
\end{align}
Although this effective coherence length is not identical to the conventional superconducting coherence length, it is closely related to the fluctuation-induced pseudogap~\cite{YANASE20031}.
As shown in Figs.~\ref{fig:T_matrix_flat}(b) and \ref{fig:T_matrix_flat}(d), the effective coherence length $\xi_{\rm eff}$ increases as the quantum metric increases.
Although the coherence length is proportional to the square root of the quantum metric in isolated flat-band superconductors that satisfy uniform pairing condition~\cite{Chen2024-ye,Hu2025-dm}, as also seen for Model~2 in Fig.~\ref{fig:T_matrix_flat}(d), the effective coherence length is not proportional to the quantum metric parameter $d$ in Model~1 because of the band touching and non-uniform pairing.

The change in the spatial correlations of superconducting fluctuations is accompanied by a pronounced modification of the single-particle spectrum.
Figures~\ref{fig:spectral_functions}(a)--\ref{fig:spectral_functions}(c) show the evolution of the single-particle spectrum with increasing the quantum metric parameter $d$ for Model~1.
In all panels, a pseudogap appears in the flat lower band: the flat band in the non-interacting Hamiltonian is split with the gap magnitude $\Delta_{\mathrm{pg}}$ [illustrated in  Fig.~\ref{fig:spectral_functions}(a)] due to the preformed Cooper pair formation.
For a zero or small $d$, a large pseudogap appears, which is progressively suppressed as the quantum metric parameter $d$ increases.
Qualitatively the same effects of quantum geometry are obtained for Model~2 as shown in Figs.~\ref{fig:spectral_functions}(d)--\ref{fig:spectral_functions}(f), where increasing the quantum metric parameter $\chi$ suppresses the pseudogap.
The consistent behavior observed in the two qualitatively different flat-band models indicates that the suppression of the fluctuation-induced pseudogap is a robust quantum-geometric effect.
For Model~1, although the transition temperature also varies with $d$ for a fixed interaction strength $U$, pseudogap suppression persists even when the interaction strength is adjusted to maintain a fixed transition temperature, as shown in Appendix~\ref{app:same_tc}.

The flat band in the absence of quantum geometry is regarded as a collection of decoupled zero-dimensional systems, giving rise to significant superconducting fluctuations that cause the pseudogap in the single-particle spectrum. 
In contrast, the quantum metric generates a finite coherence length for superconducting fluctuations, thereby suppressing the fluctuation-induced pseudogap. 
Thus, quantum geometry provides a mechanism for controlling many-body phenomena through the spatial extent of superconducting fluctuations.

\subsection{Quantum-geometry-induced crossover from non-Fermi liquid to Fermi liquid}\label{subsec:crossover}
\begin{figure}[htbp]
    \centering
    \includegraphics[width=1.0\linewidth]{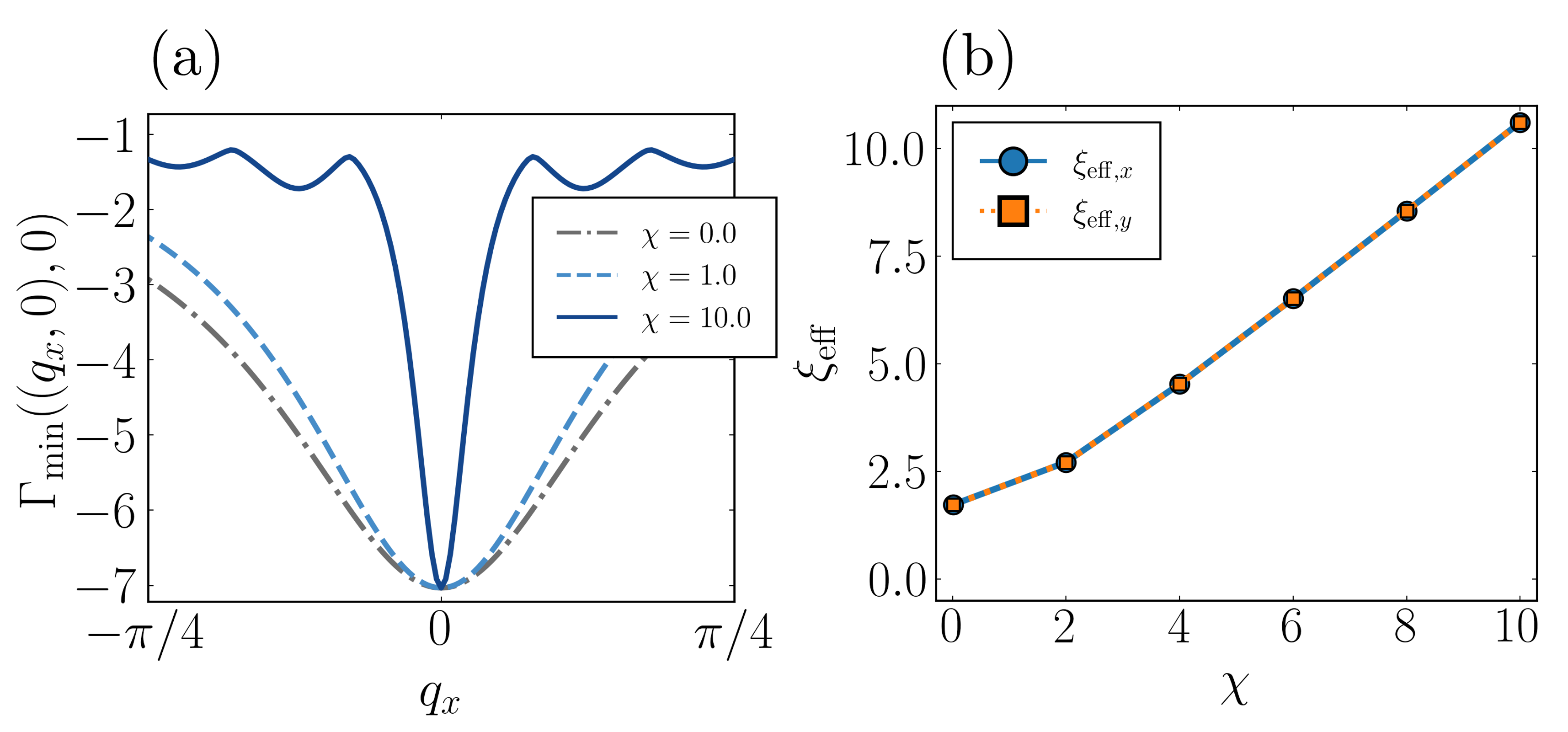}
    \caption{(a) Momentum dependence of $\Gamma_{\rm min}(\bm{q})$ for Model~2 with finite band dispersion ($w=1.0$). 
    (b) The effective coherence length $\xi_{\rm eff}$ as a function of the quantum metric parameter $\chi$. We set $U=0.6$.
    All results are evaluated at $T=1.2T_{\mathrm c}$.}
    \label{fig:T_matrix_disp}
\end{figure}
\begin{figure*}[htbp]
    \centering
    \includegraphics[width=1.0\linewidth]{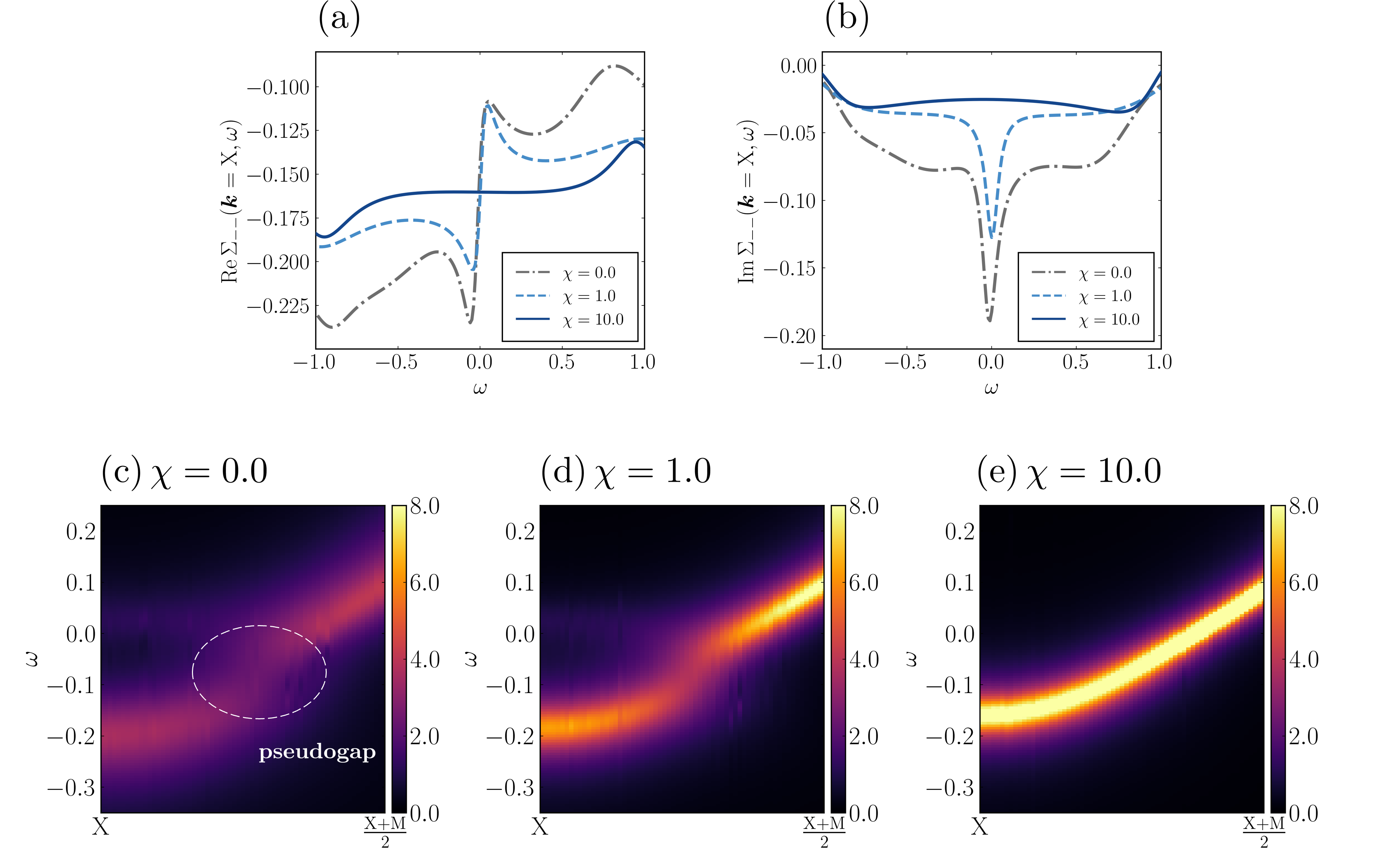}
    \caption{Quantum-geometry-induced crossover from non-Fermi-liquid to Fermi-liquid behavior in Model~2 with finite band dispersion ($w=1.0, U=0.6$). 
    (a) The real part and (b) the imaginary part of the self-energy at $\bm{k}=\mathrm{X}=(\pi, 0)$ for the lower band. 
    Intensity plots of the single-particle spectrum for (c) $\chi=0.0$, (d) $\chi=1.0$, and (e) $\chi=10.0$.
    The spectra in (c)–(e) are shown in the low-energy region along the momentum path from X to the midpoint between X and M.
    All results are evaluated at $T=1.2T_{\mathrm c}$.}
    \label{fig:crossover}
\end{figure*}

Next, we focus on the fact that even when the system possesses a finite band dispersion, quantum geometry can modify the superconducting properties.
In the following, we show that quantum geometry modulates superconducting fluctuations, and consequently induces the crossover from non-Fermi-liquid to Fermi-liquid behavior, a signature of the BCS--BEC crossover. 
Figure~\ref{fig:T_matrix_disp}(a) shows the momentum dependence of $\Gamma_{\min}(\boldsymbol{q})$ in Model~2 with finite band dispersion $w=1$.
Similarly to the flat-band system, the peak of $\Gamma_{\min}(\boldsymbol{q})$ becomes sharper as the quantum metric increases, although this peak inherently exists even without quantum-geometric effects, reflecting the finite band dispersion.
As shown in Fig.~\ref{fig:T_matrix_disp}(b), while the underlying band dispersion provides a finite effective coherence length $\xi_{\rm eff}$ even in the absence of the quantum metric, it is further enhanced as the quantum metric increases.

Accompanying this continuous increase in the effective coherence length, the self-energy and single-particle spectrum undergo dramatic changes.
Figures~\ref{fig:crossover}(a) and \ref{fig:crossover}(b) show the real and imaginary parts of the self-energy at the Fermi surface, respectively.
In the absence of the quantum metric ($\chi=0$), the self-energy exhibits pronounced features characteristic of the pseudogap: a sharp negative peak in its imaginary part and a positive slope in its real part near $\omega=0$. 
These features are in marked contrast to the behavior expected from Fermi-liquid theory. 
These non-Fermi-liquid behaviors are suppressed as the quantum metric increases.
For a large quantum metric, the low-energy self-energy progressively  recovers the Fermi-liquid-like behavior, with its imaginary part approaching a quadratic frequency dependence.

The self-energy, governed by quantum geometry, directly influences the single-particle spectrum.
Although the lower band in the non-interacting model crosses the Fermi level at the X point, the self-energy shifts the low-energy spectrum to lower energies, so that the pseudogap appears at momenta between X and M.
To highlight this feature, Figs.~\ref{fig:crossover}(c)--\ref{fig:crossover}(e) focus on the low-energy region and the momentum range from X to the midpoint of the X--M path, where the pseudogap in the single-particle spectrum is observed.
In accordance with the crossover from non-Fermi-liquid to Fermi-liquid behavior in the self-energy, Figs.~\ref{fig:crossover}(c)--\ref{fig:crossover}(e) show that the pseudogap in the single-particle spectrum is continuously suppressed and Fermi-liquid-like quasiparticles emerge as the quantum metric parameter $\chi$ increases.
We have confirmed that the self-energy at a momentum where the pseudogap appears shows similar evolution as Figs.~\ref{fig:crossover}(a) and \ref{fig:crossover}(b) as $\chi$ is varied (see Appendix~\ref{app:se_pc}).
Taken together, these changes indicate a continuous crossover from a BEC-like strong-coupling regime toward a BCS-like weak-coupling regime~\cite{Nozires1985-br,Janko1997-he,Yanase1999-dw,Rohe2001-qg,Perali2002-yv,Tsuchiya2009-vi,YANASE20031} driven by quantum geometry.

\begin{figure*}[htbp]
    \centering
    \includegraphics[width=1.0\linewidth]{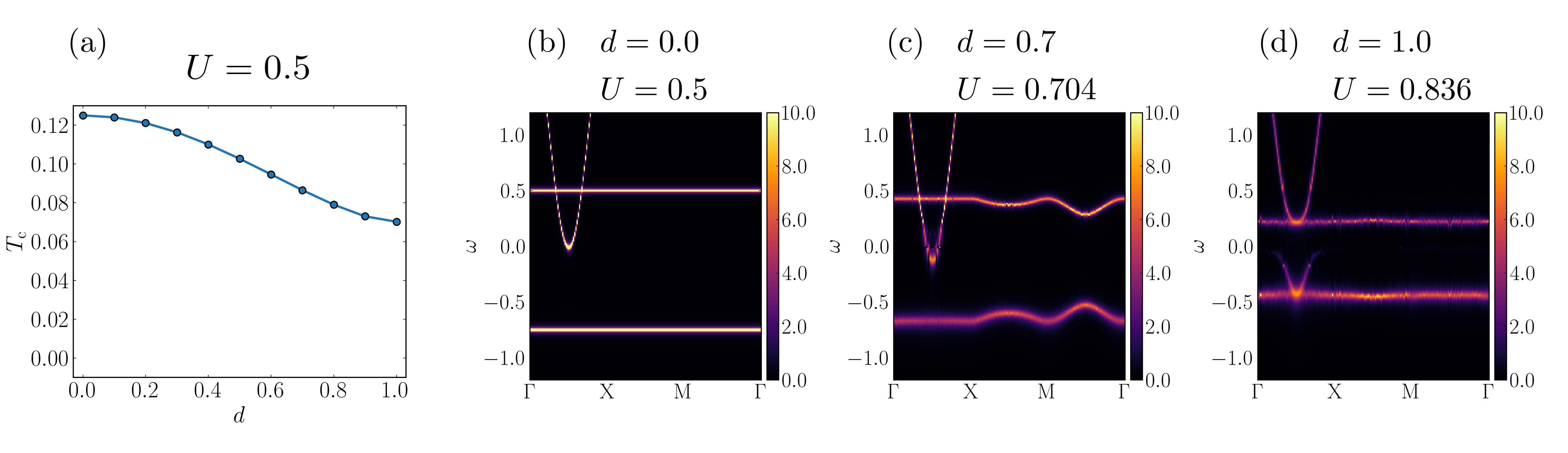}
    \caption{
    (a) The transition temperature $T_{\mathrm c}$ as a function of the quantum metric parameter $d$ at a fixed interaction strength $U=0.5$.
    Single-particle spectral functions for (b) $d=0.0$, (c) $d=0.7$, and (d) $d=1.0$ are calculated with the interaction strengths indicated on top of panels, which are adjusted to maintain the same $T_{\mathrm c}$ as for $d=0$ and $U=0.5$.
    All results are evaluated at $T=1.2T_{\mathrm c}$.}
    \label{fig:same_tc}
\end{figure*}

\section{SUMMARY AND DISCUSSION}\label{sec:Summary}
In this study, we have demonstrated that quantum geometry is a control parameter that suppresses the fluctuation-induced pseudogap and drives a continuous crossover from a non-Fermi-liquid state to a Fermi-liquid state in superconductors above $T_{\rm c}$. 
The suppression of pseudogap was demonstrated in two qualitatively distinct flat-band models: a finite-range hopping model with band-touching points and an infinite-range hopping model of isolated flat bands without band touching.
Furthermore, in Model~2 with finite band dispersion, increasing the quantum metric suppresses the pseudogap and makes the low-energy self-energy increasingly Fermi-liquid-like, indicating a continuous crossover from a BEC-like strong-coupling regime toward a BCS-like weak-coupling regime.

An interpretation of these spectral changes is provided by the quantum-geometric modification of the spatial structure of superconducting fluctuations.
As the quantum metric increases, the momentum-space peak of the T-matrix around $\bm{q}=0$ becomes sharper, corresponding to the increase in the effective coherence length $\xi_{\rm eff}$.
In the flat-band limit, the finite effective coherence length originates from quantum geometry, whereas in the dispersive-band systems the quantum metric enhances the coherence length in addition to the contribution from the band dispersion. These observations are consistent with previous calculations~\cite{Chen2024-ye,Iskin2024-xl,Hu2025-dm}. 
Taken together, our results indicate that the quantum-metric-induced increase in the spatial correlation scale of superconducting fluctuations is associated with the suppression of the fluctuation-induced pseudogap and, in dispersive systems, with the evolution toward a BCS-like weak-coupling regime.

More broadly, this work demonstrates that quantum geometry can play an essential role not only in mean-field superconducting properties but also in interaction-driven fluctuation phenomena beyond mean-field theory.
In particular, we have demonstrated that the geometry of Bloch wave functions can influence the spectral properties of interacting electron systems in the critical region.
These results may therefore be relevant to nearly-flat-band systems such as moir{\'e} superconductors~\cite{Cao2018-gn,Tanaka2025-sy}, where quantum geometry and strong superconducting fluctuations can coexist and pseudogap behavior has been experimentally observed~\cite{Oh2021-iz}.
A direct connection to specific materials, however, requires incorporating realistic multiband electronic structures, material-dependent interactions, and possible pairing symmetries beyond the simplified two-band models considered in this paper.
Extending the present framework in these directions will clarify how quantum geometry shapes superconducting fluctuations and many-body phenomena in realistic quantum materials.

\begin{acknowledgments}
We appreciate helpful discussions with Taisei Kitamura and Chang-geun Oh.
This work is supported by JSPS KAKENHI (Grant Numbers JP22H01181, JP22H04933,
JP23K17353, JP23K22452, JP24K21530, JP24H00007, JP25H01249, JP26H02016).
\end{acknowledgments}

\appendix
\section{Pseudogap suppression at a fixed $T_{\mathrm c}$}\label{app:same_tc}

In Sec.~\ref{subsec:pg_flat}, we demonstrate the suppression of the pseudogap in Model~1 by varying the quantum metric parameter $d$ at a fixed interaction strength $U=0.5$ and a fixed ratio $T/T_{\mathrm c}=1.2$.
However, as shown in Fig.~\ref{fig:same_tc}(a), the transition temperature $T_{\mathrm c}$ decreases with increasing $d$, resulting in different absolute temperatures for different values of $d$.
To examine whether the pseudogap suppression persists at a fixed transition temperature, we adjust the interaction strength $U$ for each $d$ to maintain the same $T_{\mathrm c}$ as for $d=0$ and $U=0.5$.
Figures~\ref{fig:same_tc}(b)--\ref{fig:same_tc}(d) show the resulting single-particle spectral functions for each $d$.
Although the attractive interaction becomes stronger with increasing $d$, the pseudogap is still suppressed.
This result further supports the conclusion that quantum geometry suppresses the pseudogap by modifying the momentum-space structure of superconducting fluctuations.

\section{Self-energy at pseudogap momentum}\label{app:se_pc}
\begin{figure}[htbp]
    \centering
    \includegraphics[width=1.0\linewidth]{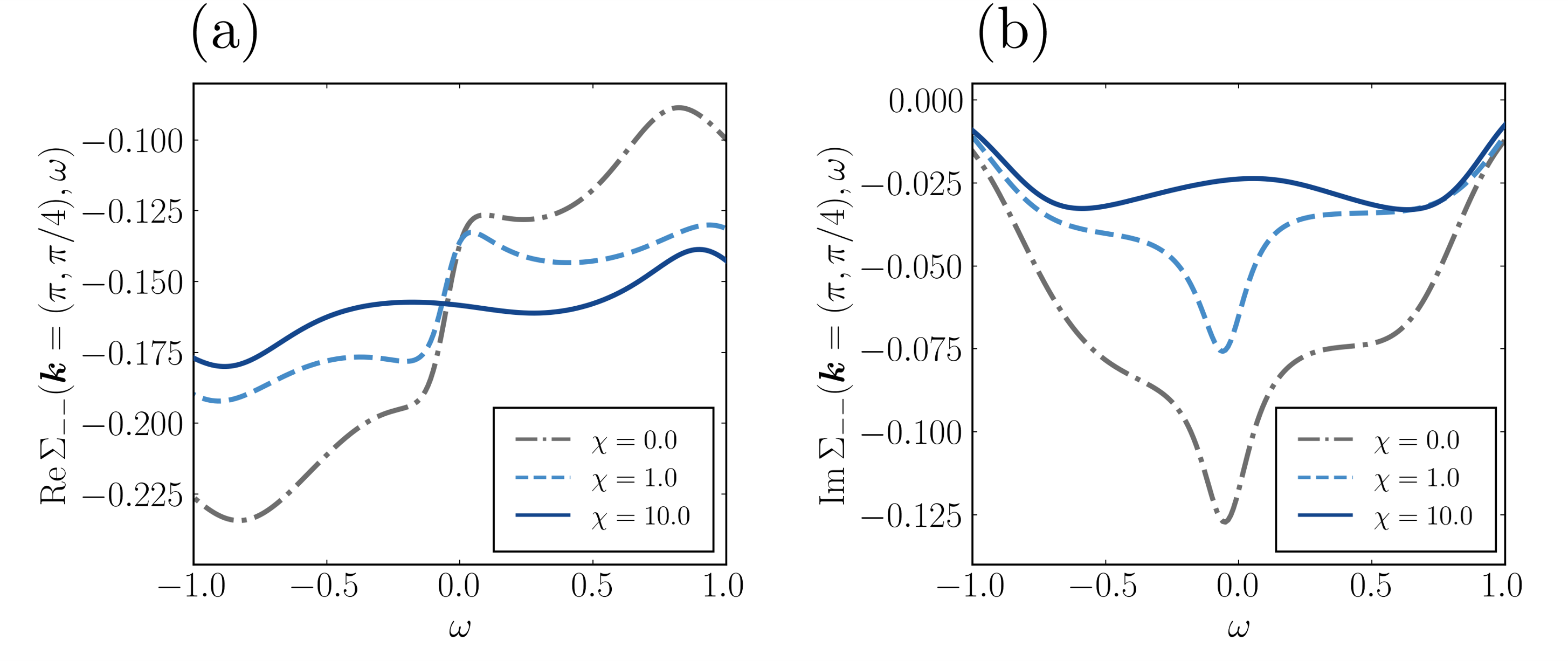}
    \caption{Self-energy at $\bm{k}=(\pi,\pi/4)$, where the pseudogap appears, for Model 2. The parameters ($w=1.0$, $U=0.6$) are the same as Fig.~\ref{fig:crossover} for the dispersive-band model.   
    (a) Real part and (b) imaginary part of the self-energy for the lower band with various quantum metric parameters $\chi$. 
    All results are evaluated at $T=1.2T_{\mathrm c}$.}
    \label{fig:se_pg}
\end{figure}
As discussed for Model~2 in Sec.~\ref{subsec:crossover}, the pseudogap in the single-particle spectrum appears away from the X point in momentum space due to the self-energy-induced shift of the low-energy band. 
We show the self-energy at $\bm{k}=(\pi,\pi/4)$, where the pseudogap appears, in Fig.~\ref{fig:se_pg}. 
For a small quantum metric parameter $\chi$, the imaginary part of self-energy exhibits a pronounced negative peak associated with pseudogap formation.
This non-Fermi liquid behavior is progressively suppressed as $\chi$ increases, consistent with the behavior of the self-energy at the X point shown in Figs.~\ref{fig:crossover}(a) and \ref{fig:crossover}(b).
These results provide further evidence for a crossover from non-Fermi-liquid to Fermi-liquid behavior accompanying the suppression of the pseudogap.

\bibliography{ref}

\end{document}